\documentclass[
 aps,
 pre,%
 amsmath,amssymb,
 reprint,
showkeys,
]{revtex4-2}

\usepackage[T1]{fontenc}
\usepackage[utf8]{inputenc}
\usepackage{amsmath}
\usepackage{amssymb}
\usepackage{graphicx}
\usepackage{refstyle}
\usepackage{multirow}
\usepackage{tabularx,booktabs}
\usepackage{multirow}
\usepackage[
  citecolor=blue,
  colorlinks,
  linkcolor=blue,
  urlcolor=blue,
]{hyperref}
\usepackage[dvipsnames]{xcolor}
\usepackage{dcolumn}% Align table columns on decimal point
\usepackage{bm}% bold math
\usepackage{soul}

\begin{document}
\title{Subduing always defecting mutants by multiplayer reactive strategies: Non-reciprocity versus generosity}
\author{Shubhadeep Sadhukhan}
\email{deep@iitk.ac.in}
\affiliation{
	Department of Physics,
	Indian Institute of Technology Kanpur,
	Uttar Pradesh 208016, India
}

\author{Ashutosh Shukla}
\email{ashus@iitk.ac.in}
\affiliation{
  Department of Physics,
  Indian Institute of Technology Kanpur,
  Uttar Pradesh 208016, India
}
\author{Sagar Chakraborty}
\email{sagarc@iitk.ac.in}
\affiliation{
  Department of Physics,
  Indian Institute of Technology Kanpur,
  Uttar Pradesh 208016, India
}

\date{\today}% It is always \today, today,
%%% Abstract~----

\begin{abstract}
A completely non-generous and reciprocal population of players can create a robust cooperating state that cannot be invaded by always defecting free riders if the interactions among players are repeated for long enough. However, strict non-generosity and strict reciprocity are ideal concepts, and may not even be desirable sometimes. Therefore, to what extent generosity or non-reciprocity can be allowed while still not be swamped  by the mutants, is a natural question. In this paper, we not only ask this question but furthermore ask how generosity comparatively fares against non-reciprocity in this context. For mathematical concreteness, we work within the framework of multiplayer repeated prisoner's dilemma game with reactive strategies in a finite and an infinite population; and explore the aforementioned questions through the effects of the benefit to cost ratio, the interaction group size, and the population size.
\end{abstract}
%%%%%%%%%%%%%%%%
\maketitle
%%%%%%%%%%%%%%%%%%%%%
\section{Introduction}\label{sec:intro}
Cooperation~\cite{1971_Trivers,1982_Smith, 2006_Axelrod, 2006_Nowak,2009_Gintis,2017_PJRWS_PR} is a primary tenet of the social structure of many pro-social beings, and plays a major role in the sustenance of various socio-economic and biological systems~\cite{1998_Smith, 2013_N, 2011_Bourke}. Since Darwinism suggests that a defector would reap the comparable benefit as cooperators while incurring no cost unlike the cooperators, it is an intriguing question why one should cooperate even when being selfish is more advantageous. Focussed attempts to find answers to this question have lead to the understanding that the cooperation can emerge via direct reciprocity~\cite{1971_Trivers, 2008_PTON_JTB}, indirect reciprocity~\cite{Nowak2005}, kin selection~\cite{HAMILTON_1964}, network reciprocity~\cite{2006_OHLN_Nature}, group selection~\cite{SMITH1964}, voluntary participation~\cite{2002_Szabo_PRL}, punishment~\cite{Chen2015}, and a few other mechanisms like generosity~\cite{Kurokawa2010,2017_park_pre,kurokawa_2019}. Evolutionary game theory~\cite{SP1973,smith1982book} is the well-exploited formalism for examining such ideas; and the prisoner's dilemma game~\cite{1965_RC}, which epitomizes the problem of the evolution of cooperation, is the simplest nontrivial testbed for investigating these ideas.

Among the solution concepts of non-cooperative games, the Nash equilibrium~\cite{1950_N_PNAS}---a strategy set such that no player can increase her payoff by unilaterally deviating from her own strategy---is arguably the most well-known. A refinement of the Nash equilibrium in the context of evolutionary games is evolutionarily stable strategy (ESS)~\cite{1982_Smith} which is of foremost interest to evolutionary game theorists. The ESS is a strategy adopted by all the individuals of an infinitely large population such that the host population can not be invaded by an infinitesimal fraction of mutants. In the context of the problem of cooperation in the prisoner's dilemma, defection happens to be an ESS that cannot be invaded by cooperators. However, in a finite population, where there is inherent stochasticity, there is finite probability that few  cooperators can take over the host population of defectors. Thus, in a bid to contextualize the ESS for finite population, an evolutionary stable strategy (ESS$_N$) in finite population (of size $N$) was proposed~\cite{Nowak2004}: A strategy $S_1$ to said be ESS$_N$ if, given a mutant strategy $S_2$, (i) selection opposes $S_2$ to invade $S_1$ which means a single mutant $S_2$ in a population of $S_1$ has a lower fitness, and (ii) selection opposes $S_2$ to replace $S_1$ that means the fixation probability of $S_2$---the probability with which a single mutant $S_2$ becomes the ancestor of all the individuals in the eventual population---is less than $1/{ N}$.

Of course, one-shot games, whether in infinite or finite population, do not allow for either reciprocity or generosity. They can be realized in the setting of repeated games because a player gets opportunity to respond---either reciprocally or generously---to the action of the other participants. In repeated games, the shadow of the future~\cite{skyrms2003book} looms large: The probability of a subsequent interaction in a repeated game need not be unity but some $\delta$ between zero and one. In such cases, one can equivalently think that the payoff in every subsequent interaction gets discounted by the multiplicative factor $\delta$. In the framework of repeated games, the strategy set is enormous since there can be many sequences of possible actions for each player. However, in rather general setting of evolutionary games, it is desirable that the players are not burdened with the requirements of rationality and high cognitive load. Consequently, it makes pragmatic sense to keep focus on a much smaller set of strategies, viz., reactive strategies: A player's reactive strategy is a sequence of actions such that each action played at a step is merely a reaction to what the opponent played in the immediately preceding step. Some of these strategies---the most famous one being the tit-for-tat strategy (TFT)~\cite{2006_Axelrod}---can actually render cooperation evolutionarily stable; when enough repeated interactions happen, the short term advantage of defecting mutants against TFT is not worthy over the long term loss. 

However, when an entire population is under consideration and not just two isolated player, multi-agent interactions are very ubiquitous and natural~\cite{Joshi1987,Boyd1988}. Naturally, the paradigmatic role of the prisoner's dilemma is taken over by the $n$-player prisoner's dilemma game~\cite{Kurokawa2010, Kurokawa2018} and public goods game~\cite{2009_WNH_PNAS, 2012_Wang_JStat} for the multiplayer interactions. Apparently, as the size of reciprocating groups involved in multiplayer interaction increase, the emergence of reciprocal cooperation becomes harder~\cite{Boyd1988}. Typically, cooperators punish the defectors by withholding further cooperation in a two-player interaction; in contrast, in the case of multiplayer social interaction, other cooperating peers may suffer due to the withheld future cooperation from a cooperator. In such cases, an optimal generosity turns out to be the better option in the emergence of cooperation~\cite{Kurokawa2010}. Mismatch of the intent and the actual outcome due to the mistakes are common. In such cases, generosity may play a crucial role in the emergence of the cooperation in multiplayer repeated interaction. Generous individuals provide relatively more opportunities for others to cooperate.

The question that we ask in this paper in the backdrop of the emergence of cooperation through multiplayer reactive strategies is: how much the individuals in the population can afford to be non-reciprocative and generous and still not be invaded by defectors? It is well-known that extremely generous populations would be easily exploited by defectors and so would be a highly non-reciprocal population. Hence the related question: how does the maximum extent of generosity that an evolutionary stable population consisting of completely reciprocal individuals can sustain fare against the maximum extent of non-reciprocity that an evolutionary stable population consisting of completely non-generous individuals can sustain? Furthermore, what are the relative effects of cost, benefit, interaction group size, and overall population size on this comparison? Since the words, reciprocity and generosity, can be interpreted and mathematically modelled in various different ways, we first describe our setup precisely in the context of multiplayer repeated prisoner's dilemma in the next section so that we can address the aforementioned questions unambiguously.
%%%%%%%%%%%%%%%%%%%%%%%%%%%%%%%%%%%%%%%%%%%%%%5
\section{The Setup}\label{sec:setup}
To begin with, let us consider a group of $n$ players involved in one-shot multiplayer prison's dilemma game described as follows. Each player can play one of the two actions---cooperate or defect. All the players benefit from the cooperators' contributions but only the cooperators pay cost. The payoff of a cooperating player, when there are $k$ cooperators (and $n-k$ defectors) present, is defined to be $bk/n-c$ and that of a defector is $b k/n$, where the positive real numbers, $b$ and $c$, represent benefit and cost respectively. Thus, if there were only cooperators, the payoff of each player would be $b-c$; and hence if a cooperator changes to a defector then she gets $b-b/n$. Similarly, if there were only defectors, the payoff of each player would be $0$; and so if a defector changes to a cooperator then she gets $b/n-c$. Consequently, for the standard prisoner's dilemma to appear, we require to impose the conditions that $b-c>0$, $b-b/n>b-c$, and $0>b/n-c$; the conditions can be collectively written as $b>c>b/n>0$.

Now to extend, we create a repeated game based on this one-shot game, we assume that the one-shot game is repeatedly played for many rounds such that a player in  the group can choose her action in each round based on the past actions of the rest of the players she is interacting with. In the simplest analytically tractable setting, we assume that the focal player's action in a round depends exclusively on the actions of the other players in the immediately preceding round. In other words, the focal player's strategy---the complete ordered sequence of actions over the rounds---is a reactive strategy. Of course, the standard two-player reactive strategies need to be redefined, e.g., TFT can be generalized~\cite{Boyd1988} to define TFT$_m$ that cooperates if more than $m$ number of cooperators were present in the immediate preceding step. TFT$_m$ is called softer or harder than another TFT$_{m'}$ if $m$ is respectively smaller or greater than $m'$. 

We are interested in rather general reactive strategy, $E_m(p,q)$, with $p$ and $q$ simultaneously not zero and with cooperation at the very first step: The player with strategy $E_m(p,q)$ cooperates with a probability $p$ when there were at least $m$ cooperators in the previous round and she cooperates with a probability $q$ when the number of cooperators was less than $m$ in the previous step. Obviously, the probability $1-p$ is a measure of non-reciprocity, whereas the probability $q$ is a measure of generosity or forgiveness. A slight ambiguity creeps in when one notes that the softer version of a reactive strategy may be seen as more generous than its harder version. Partially with a view to bypassing this ambiguity we set $m=n-1$ henceforth so that in this paper, level of generosity is defined exclusively through $q$. In this context we also point out that it was shown that TFT$_{m}$ can be ESS$_N$ against always defect (ALLD) strategy only for $m=n-1$~\cite{Joshi1987,Kurokawa_2009}. Hence, given the enormity of the number of possible reactive strategies, it is pragmatic for us to selectively focus on how much less reciprocal ($p<1$) and more generous ($q>0$) the reactive strategies can be compared to the TFT$_{n-1}$ (or, $E_{n-1}(p=1,q=0)$) and still be evolutionarily stable.

The question we are asking requires us to envisage a host population of players, all with $E_{n-1}(p,q)$ reactive strategy, playing repeated multiplayer prisoner's dilemma game. Any $n$ individuals are randomly grouped for the necessary multiplayer interactions. We want to find if the host population resists being taken over by the ALLD strategy when some of the players mutate to adopt the ALLD strategy. Naturally, to this end, we require the details of the payoffs as the interactions occur. This is presented in Table~\ref{tab:payoffm} where the payoff elements $a_i$ and $b_i$  represent the payoff of a player playing ALLD and $E_{n-1}(p,q)$ respectively if $n-i$ opponents play ALLD. The next task is to find the explicit expressions for the payoff elements in terms of the parameters of the model, viz., $b$, $c$, $n$ and $\delta$. We recall, as mentioned in Sec.~\ref{sec:intro}, $\delta$---also called the discount factor---is the extra multiplicative factor by which payoff at a round is multiplied (discounted) with respect to the payoff of the immediately preceding round in the light of the shadow of the future. More precisely, the payoff at $l$th round is multiplied with $\delta^{l-1}$ to generate the effective payoff. Equivalently, $1/(1-\delta)$ is known as the expected game length which is larger when $\delta$ is larger.
\begin{table}[h!]
\begin{center}
\begin{tabularx}{240pt}{c|cccccc}
    \hline
    \multirow{2}{6em}{Strategy of focal player} & \multicolumn{6}{c}{No. of ALLD players in $n-1$ opponents} \\
        \cline{2-7}
        \rule{0pt}{3ex} & $n-1$ & $n-2$ &\ldots & $n-i$ & \ldots & $0$\\
         \hline
         ALLD & $a_1$ & $a_2$ &\ldots & $a_i$ & \ldots  & $a_n$\\
         $E_{n-1}(p, q)$ & $b_1$ & $b_2$ & \ldots  & $b_i$ & \ldots & $b_n$\\
         \hline
\end{tabularx}
\end{center}
\caption{Payoff matrix for two strategies ALLD and $E_{n-1}(p,q)$ with an underlying multiplayer prisoner's dilemma game.}
\label{tab:payoffm}
\end{table}

\begin{figure*}
\includegraphics[scale=1.15]{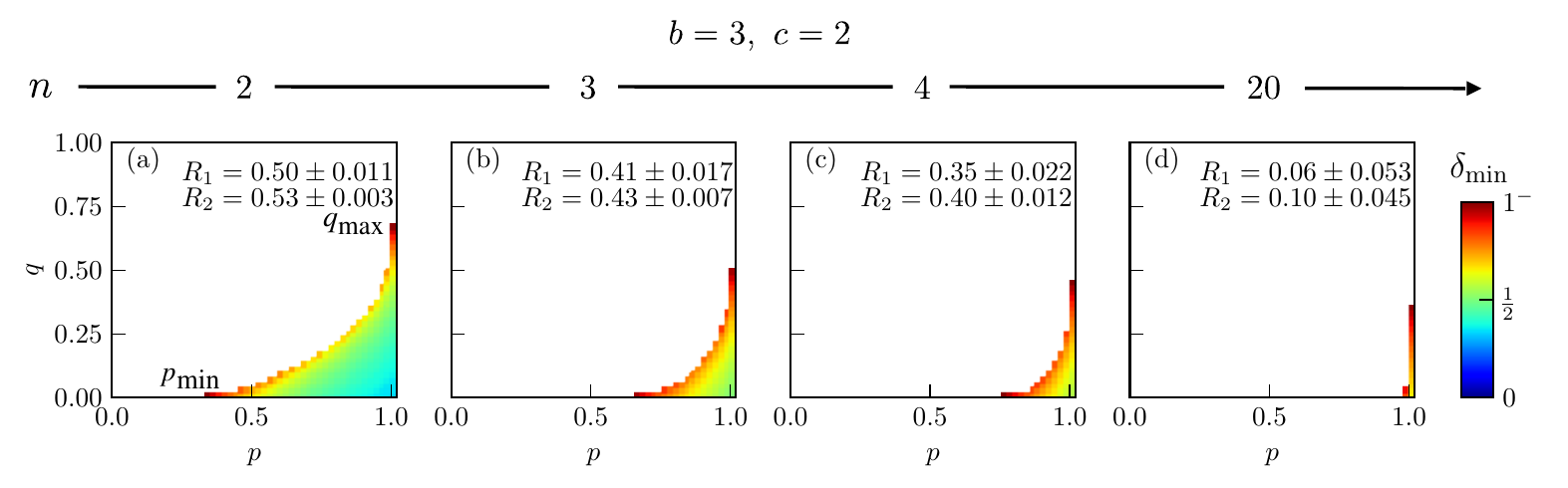}
\caption{{The number of reactive strategies, $E_{n-1}(p,q)$, that are ESS shrinks with increase in group size $n$, and $R_1$ and $R_2$ decrease:} The coloured area on the $(p,q)$ strategy space denotes the region where $E_{n-1}(p,q)$ is ESS. The variation in the colour indicates the minimum repetition probability $\delta_{\rm min}$. The benefit  and cost are set to $b=3$ and $c=2$. In subfigures (a)--(d) we set the interaction group size $n=2, ~3, ~4, ~{\rm and}~20$ respectively.}
\label{fig:normalESS}
\end{figure*}

\section{Calculation of The Payoff Elements}
\label{sec:conds}
First to calculate $a_i$, consider the situation of a group of $n$ players where, in addition to one focal player, $n-i$ number players play ALLD. Hence, remaining $i-1$ individuals play $E_{n-1}(p,q)$. At the first step, by definition, all the players with $E_{n-1}(p,q)$ cooperate. Hence, the number of cooperators is $i-1$ in the very first round and the payoff of the focal ALLD individual is ${A_1}=b(i-1)/n$. The individuals playing $E_{n-1}(p,q)$ cooperates with a probability $q$ in each subsequent round because the number of cooperators is always less than $n-1$. Therefore the probability of there being exactly $r$ cooperators in any round, $l$, after the first is $P_l(r)= {}^{i-1}C_{r}q^r(1-q)^{i-1-r}$. The expected payoff to the focal ALLD player in round $l$ is 
\begin{align}
   {A_l}=& \sum_{r=0}^{i-1} \frac{br}{n}P_l(r)= \dfrac{b(i-1)q}{n}; \quad(l>1).
\end{align}
Therefore the total expected payoff is given by,
\begin{equation}
    a_i=\sum_{l=1}^\infty {A_l}\delta^{l-1}= \frac{b(i-1)}{n} \left [ 1+ \frac{\delta q}{1-\delta}\right  ].\label{eq:ai}
\end{equation}

Next, we calculate $b_i$ for all $i\ne n$. Since there is at least one defector among the opponents of the focal player playing $E_{n-1}(p,q)$, the focal player cooperate with probability $q$ in each round after first. So, considering the case where $i$ players play $E_{n-1}(p,q)$, the probability of there being exactly $r$ cooperators in round $l$ is $P_l(r)={}^{i}C_{r}q^r(1-q)^{i-r}$. Hence, the expected payoff of focal player in round, $l$, after the first is 
\begin{align}
    {B_l}=\sum_{r=0}^i \left[ \frac{br}{n} \left(1-\frac{r}{i}\right) + 
   \left(\frac{br}{n}-c \right )\frac{r}{i}  \right]P_l(r); \quad(l>1).
   \end{align}
Here, first term inside the square brackets is due to payoff received when the focal player is not one of the $r$ cooperators and second term is when she is one of the cooperators. We also have ${B_1}=bi/n-c$. Therefore,
\begin{align}
    b_i=\sum_{l=1}^\infty {B_l} \delta^{l-1}= \left(\frac{bi}{n}-c \right)+ \left(\frac{b}{n}-\frac{c}{i} \right)\frac{\delta iq}{1-\delta}; \quad(i\ne n)  \label{eq:bi}.
\end{align}

Finally, we proceed to find $b_n$. To this end, we have to think a situation where all the $n$ players play $E_{n-1}(p,q)$. In first round, all of them cooperate so $B_1=b-c$. But at any arbitrary round (other than the first), the player cooperate either with probability $q$ or $p$ depending on how many opponents cooperated in the immediately preceding round. In other words, the probability $P_l(r)$ for having $r$ cooperators in round $l>1$ depends on $P_{l-1}(j)$ where $j$ is an integer lying between $0$ to $n$. Now, we consider different cases depending on the values of $j$ takes different values:
\begin{enumerate}
\item[(i)] When $0\leq j \leq n-2$, all the $E_{n-1}{(p,q)}$ cooperate with a probability $q$ in round $l$; for which, consequently, $P_l(r)=\alpha_rP_{l-1}(j)$ where $\alpha_r\equiv{}^nC_r q^r(1-q)^{n-r}$.

\item[(ii)] When $j=n-1$, it means there is one (focal) player who defected in round $l-1$. That player must cooperate in round $l$ with a probability $p$ as she faces $n-1$ cooperators in the $(l-1)$-th round. On the other hand, rest of the $n-1$ cooperators cooperate with a probability $q$ because they face $n-2$ cooperators in the $(l-1)$th round. Hence, the $P_l(r)$ due to the $n-1$ cooperator in the $(l-1)$th round is $P_l(r)=\beta_rP_{l-1}(n-1)$ where $\beta_r\equiv {}^{n-1}C_{r}q^r(1-q)^{n-1-r}(1-p)+{}^{n-1}C_{r-1}q^{r-1} (1-q)^{n-r}p.$ Here, the first and the second terms in square bracket respectively correspond to the cases when the focal player is not one of $r$ cooperators in round $l$ and the second term is one of $r$ cooperators in round $l$.

\item[(iii)] When $j=n$, it means each of the $n$ cooperators in round $l-1$ play with $n-1$ cooperating opponents. Therefore, each cooperator must cooperate in the round $l$ with a probability $p$. Hence, $P_l(r)=\gamma_rP_{l-1}(n)$ where $\gamma_r\equiv{}^{n}C_{r}~p^r (1-p)^{n-r}$.
\end{enumerate}

As a result, any $P_l(r)$ in terms of $P_{l-1}(j)$'s using the formalism of Markov chain: We write $\boldsymbol{\pi}_l={\sf T}\boldsymbol{\pi}_{l-1}$ where the column-vector $\boldsymbol{\pi}_l\equiv[P_l(0)\,P_l(1)\,\cdots\,P_l(n)]^T$ and the elements of the transition matrix ${\sf T}$ are given by
\begin{equation}
    t_{ij}= 
    \begin{cases}
    \alpha_i & \text{if}\ 0\leq j \leq n-2, \\
    \beta_i & \text{if}\ j=n-1,\\
    \gamma_i & \text{if}\ j=n.
    \end{cases}
\end{equation}
We note that, by definition, $\boldsymbol{\pi}_1=[0\,0\,\cdots\,0\,1]^T$

Since in the round $l>1$, the focal player with strategy $E_{n-1}(p,q)$ can either be a cooperator among the $r$ cooperating individuals with a probability ${r}/{n}$ or be a defector with a probability $1-{r}/{n}$, the payoff to the focal player in round $l$ is
  \begin{eqnarray}
B_l&=&\sum_{r=0}^n \bigg[\bigg(\frac{br}{n}-c\bigg)\frac{r}{n}+\frac{br}{n}\bigg(1-\frac{r}{n}\bigg)\bigg]P_l(r)\nonumber\\
 &=& \frac{b-c}{n}\sum_{r=0}^{n}  r[{\sf T}^l \boldsymbol{\pi}_1]_r; \quad(l>1).
 \end{eqnarray}
 Here, subscript $r$ denotes the corresponding vector element. Therefore, the total expected payoff is given by,
\begin{eqnarray}
 b_n  =\sum_{l=1}^\infty {B_l} \delta^{l-1} &=&\frac{b-c}{n} \sum_{r=0}^n r [({\sf I}- {\sf T}\delta)^{-1} \boldsymbol{\pi}_1]_r \label{eq:bn},   
\end{eqnarray}
where ${\sf I}$ is $n+1$ dimensional identity matrix. 

We may observe that for $n=2$, the expressions of $a_i$'s (Eq.~(\ref{eq:ai})) and $b_i$'s (Eq.~(\ref{eq:bi}) and Eq.~(\ref{eq:bn})) obtained here match with the analogous known analytical results in the two-player case~\cite{nowaksigmund}. However, for arbitrary $n$, their---to be precise, $b_n$'s---calculation is cumbersome; and we are able to find it only numerically, mainly because of the presence of the term, $({\sf I}- {\sf T}\delta)^{-1}$, that involves finding inverse of matrix of size $(n+1) \times (n+1)$.
%%%%%%%%%%%%%%%%

\begin{figure*}
\includegraphics[scale=1]{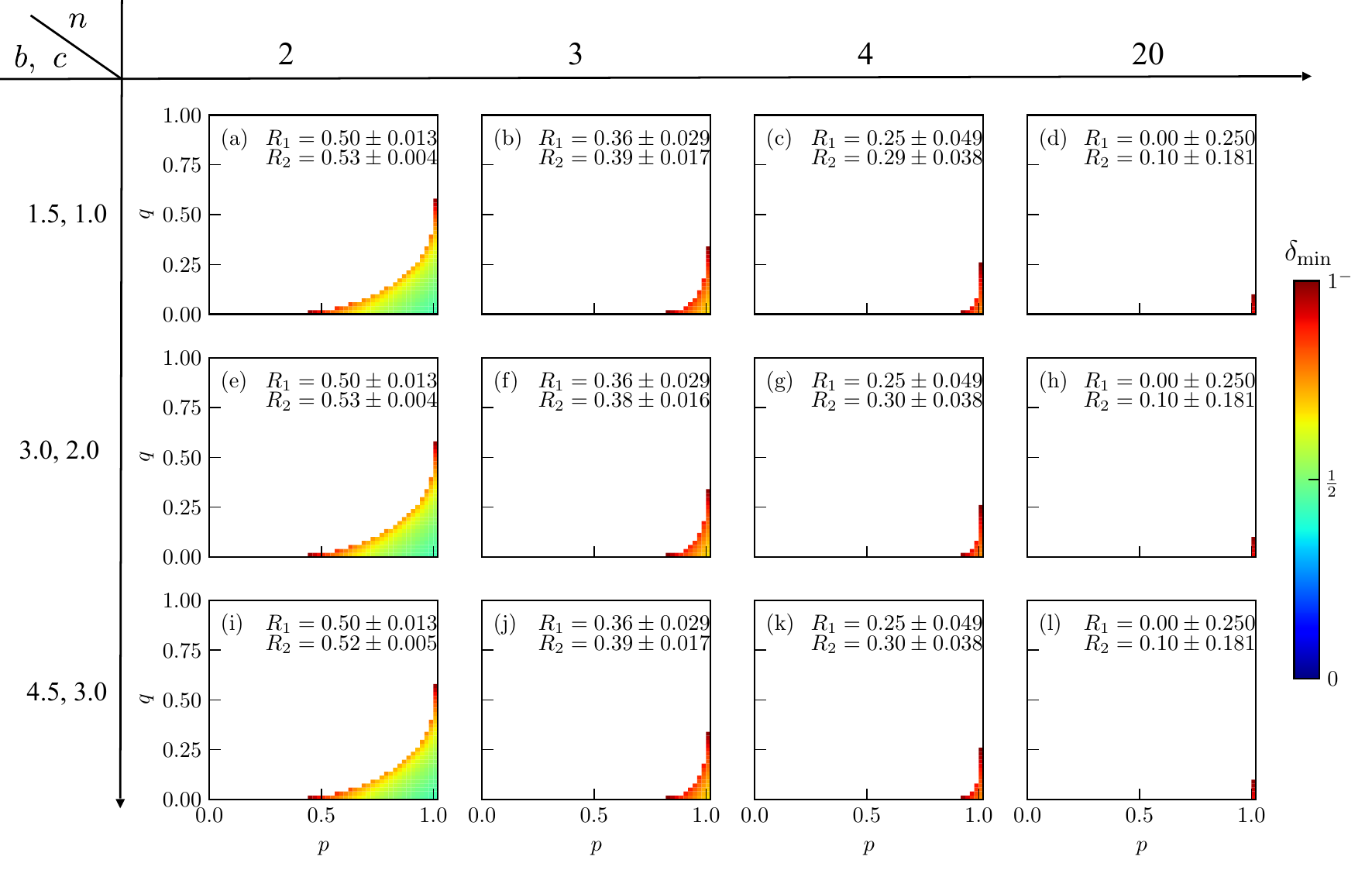}
\caption{{At fixed $b/c$, the number of reactive strategies, $E_{n-1}(p,q)$, that are ESS$_N$ shrinks with increase in group size $n$, and $R_1$ and $R_2$ decrease:} The coloured area on the $(p,q)$ strategy space denotes the region where $E_{n-1}(p,q)$ is ESS. The variation in the colour indicates the minimum repetition probability $\delta_{\rm min}$. Along each row, the interaction group size increase as $n=2, ~3, ~4, ~{\rm and}~20$. The benefit to cost ratio $b/c$ is fixed at $1.5$ by using three different parameter sets: $b=1.5,~c=1.0$; $b=3.0,~c=2.0$;  and $b=4.5,~c=3.0$. They are presented in three different rows.}
\label{fig:b_by_c}
\end{figure*}

\begin{figure*}
\includegraphics[scale=1]{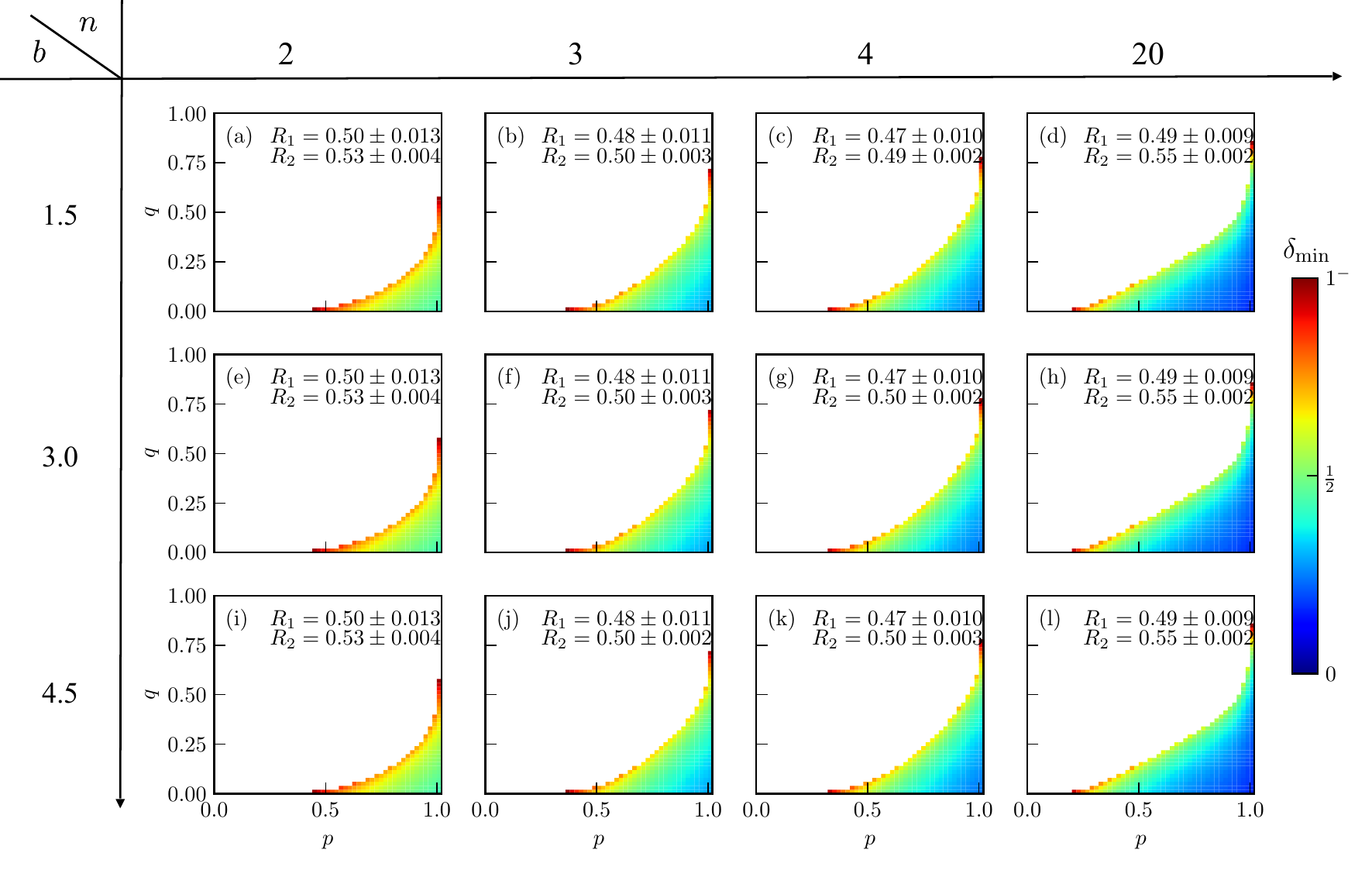}
\caption{At fixed $b/nc$, the number of reactive strategies, $E_{n-1}(p,q)$, that are ESS$_N$ increase with increase in group size $n$, and $R_1$ and $R_2$ remain $0.5$ always: The coloured area on the $(p,q)$ strategy space denotes the region where $E_{n-1}(p,q)$ is ESS. The variation in the colour indicates the minimum repetition probability $\delta_{\rm min}$. Along each row, the interaction group size increase as $n=2, ~3, ~4, ~{\rm and}~20$. The benefit to cost ratio $b/nc$ is fixed at $0.75$ while varying $b$ and $c$: $b=1.5,~3,~{\rm and}~4.5$ (and the cost $c$ is adjusted such that the ratio $b/nc$ is constant) which are respectively presented in three consecutive rows.}
\label{fig:b_by_nc}
\end{figure*}

\begin{figure}
\centering
\includegraphics[scale=0.8]{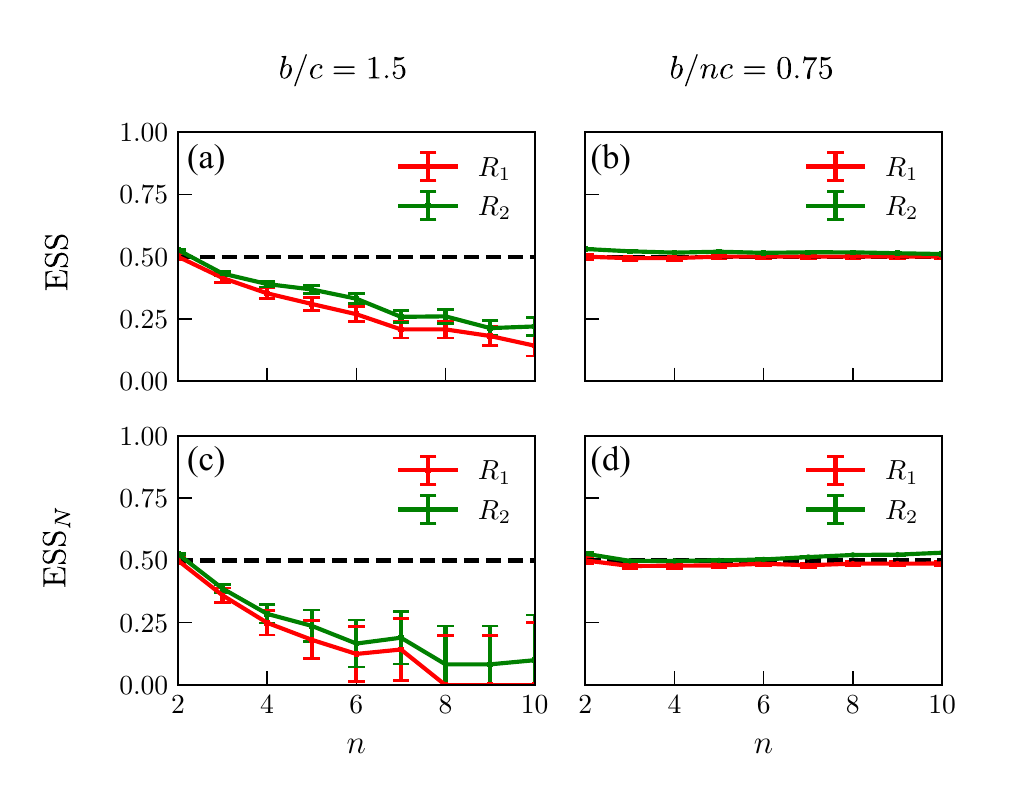}
\caption{Non-reciprocity--generosity asymmetry's dependence on benefit to cost ratio: $R_1$ (red line) and $R_2$ (green line) are plotted against the interaction group size $n$ in infinite (subplots (a) and (b)) and finite but large (subplots (c) and (d)) populations. $b/c$ is fixed at 1.5 for subplots (a) and (c). $b/nc$ is fixed to 0.75 for subplots (b) and (d).}
\label{fig:asymmetry}
\end{figure}

\begin{figure*}
\includegraphics[scale=1]{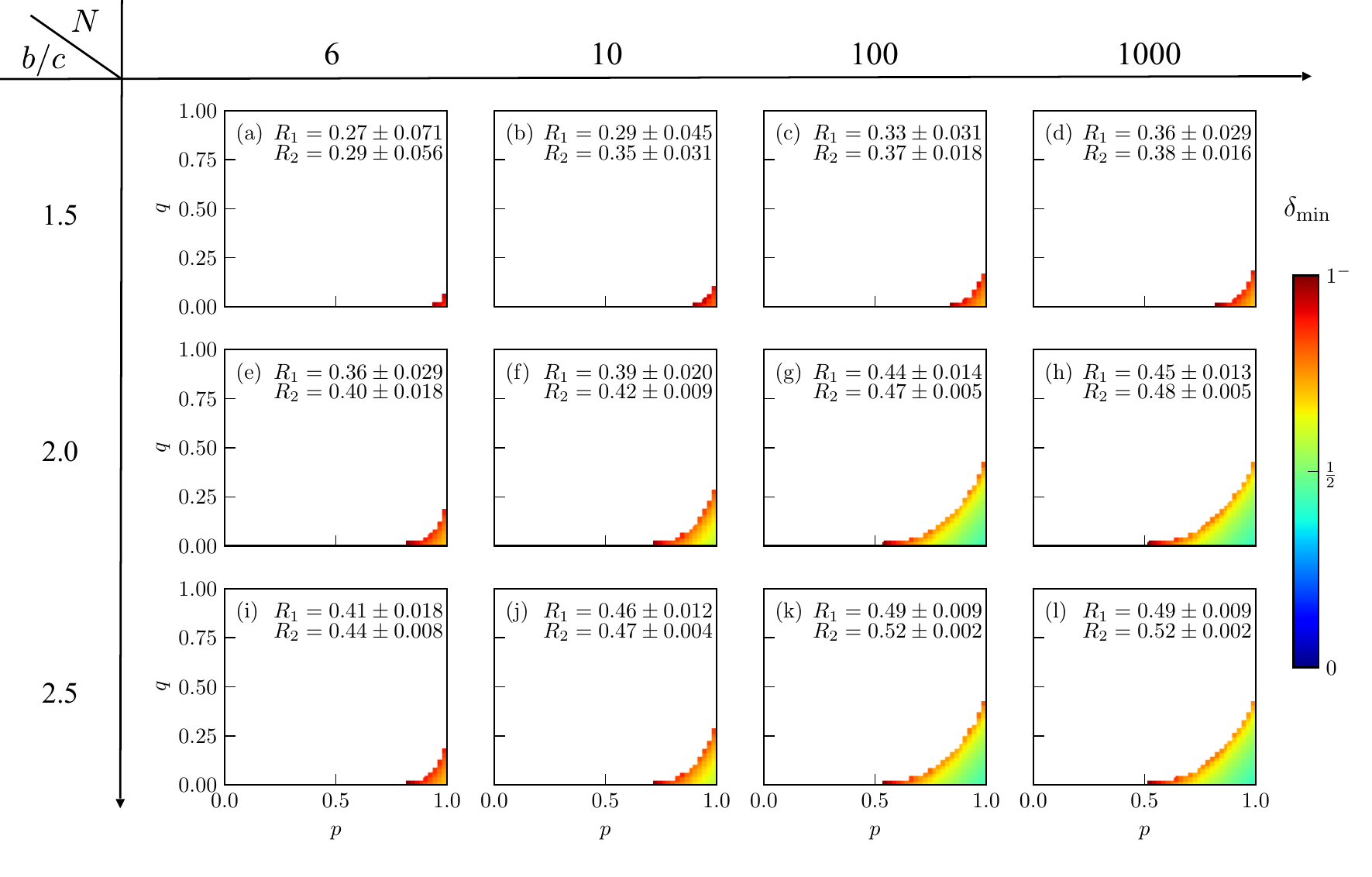}
\caption{At fixed group size, the number of reactive strategies, $E_{n-1}(p,q)$, that are ESS$_N$ increase with increase in population size $N$, and $R_1$ and $R_2$ increase: The coloured area on the $(p,q)$ strategy space denotes the region where $E_{n-1}(p,q)$ is ESS. The variation in the colour indicates the minimum repetition probability $\delta_{\rm min}$. Along each row, the population size increase as $N=6, ~10, ~100, ~{\rm and}~1000$. We set the group size of interaction $n=3$. We vary $b/c$: $b/c=1.5,~2,~{\rm and}~2.5$ which are respectively presented in three consecutive rows.}
\label{fig:finiteN}
\end{figure*}
%%%%%%%%%%%%%%%%%%%%%%%%%%%%%%%%%
\section{Infinite Population: ESS}
\label{sec:ess}
Now let us first consider the case of an infinite population where individuals play infinitely repeated multiplayer prisoner dilemma in randomly formed groups of size $n$; the strategy of every player is the same $E_{n-1}(p,q)$. We want to find out those values of $p$ and $q$ for which the population is evolutionarily stable against ALLD mutants. In other words, we want to find those $E_{n-1}(p,q)$'s that are ESS against ALLD.

The marginal payoff for a ALLD strategist is $U_{A}=\sum_{k=1}^{n}{}^{n-1}C_{n-k}x^{n-k}(1-x)^{k-1}a_k$, while that of a $E_{n-1}(p,q)$ strategist is $U_{E}=\sum_{k=1}^{n}{}^{n-1}C_{n-k}x^{n-k}(1-x)^{k-1}b_k$. Hence, the host population can resist the invasion by an infinitesimal (i.e., $x\to0$) fraction of ALLD mutant if  $U_{E}>U_{A}$, i.e., if $b_n>a_n$. Similar argument leads to the conclusion that the condition for the ALLD to be ESS is $a_1>b_1$ which is always satisfied as can be seen from 
Eq.~(\ref{eq:ai}) and Eq.~(\ref{eq:bi}), and the fact that $c>b/n$. 

The condition, $a_{n}<b_n$, needed for $E_{n-1}(p,q)$ to be ESS has to be numerically found using Eq.~(\ref{eq:ai}) and Eq.~(\ref{eq:bn}). This depends on $b$, $c$, $n$, and $\delta$. We fix $b$ and $c$ to 3 and 2 respectively, and find the minimum discount factor $\delta=\delta_{\rm min}$ at which $E_{n-1}(p,q)$ becomes ESS for different values of $n$. In Fig.~\ref{fig:normalESS}, we present all the $E_{n-1}(p, q)$'s that are ESS. We immediately note that when $n=2$, the maximum extent of generosity ($q=q_{\rm max}$) that an evolutionary stable population consisting of completely reciprocal individuals ($E_{n-1}(1,q)$) can sustain is equal to the maximum extent of non-reciprocity ($1-p=1-p_{\rm min}$) that an evolutionary stable population consisting of completely non-generous individuals ($E_{n-1}(p,0)$) can sustain. However, as the group size $n$ increases, the difference $q_{\rm max}-(1-p_{\rm min})$ increases as well. In other words, the sustenance of cooperation (that, of course, coexists with defection in general) becomes relatively harder for non-reciprocal reactive strategies compared to the generous reactive strategies as the group size increases. 

Since tracking this asymmetry between maximal generosity and maximal non-reciprocity is the main goal of this paper, we precisely quantify this asymmetry using two parameters, $R_1$ and $R_2$, which lie between zero to one. First one is simpler:
\begin{equation}
R_1\equiv\frac{1-p_{\rm min}}{q_{\rm max}+1-p_{\rm min}}.
\end{equation}
A value of 0.5 for $R_1$ implies symmetry; lesser than half means $q_{\rm max}>(1-p_{\rm min})$, while greater than half means $q_{\rm max}<(1-p_{\rm min})$. This parameter only compares $E_{n-1}(1,q_{\rm max})$ with $E_{n-1}(p_{\rm min},0)$, but there are infinitely many strategies which are ESS. Thus, in oder to measure the asymmetry about the line $q=1-p$ in the $p$-$q$ strategy space, we need to include all such strategies. This brings us to the next parameter:
\begin{equation}
R_2\equiv\frac{1}{\#(\Delta)}\sum_{\Delta} \frac{\#(q=1-p-\Delta,q=1-p)}{\#(q=1-p-\Delta,q=1-p+\Delta)}.
\end{equation}
Here the number of ESS strategies, $E_{n-1}(p,q)$, between $q=1-p-\Delta$ and $q=1-p$ has been denoted by $\#(q=1-p-\Delta,q=1-p)$; the number of ESS strategies, $E_{n-1}(p,q)$, between $q=1-p-\Delta$ and $q=1-p+\Delta$ has been denoted by $\#(q=1-p-\Delta,q=1-p+\Delta)$; the sum is over many randomly chosen values of $\Delta$; and the total number of values of $\Delta$ chosen is denoted by $\#(\Delta)$. In this paper, we have taken $\#(\Delta)=10$. Again one note that $R_2=0.5$ implies complete symmetry about the line $q=1-p$. If $R_2 > 0.5$ that means more strategies are evolutionarily stable below the line $q=1-p$ (i.e., relatively more non-reciprocal strategies). Similarly, $R_2<0.5$ that means more strategies are stable above the line $q=1-p$ (i.e., relatively more generous strategies). Going back to 
Fig.~\ref{fig:normalESS}, we observe how $R_1$ and $R_2$ decrease with increase in $n$ meaning, as concluded pictorially earlier that the sustenance of cooperation becomes relatively easier for generous reactive strategies compared to the non-reciprocal reactive strategies as the group size increases. 

\section{Finite population: ESS$_N$}
All real populations are finite and hence the effect of drift~\cite{Kimura1968, Kimura1983} is unavoidable. In fact, whatever kind of host population be there, even a single mutant individual has finite probability of taking over the entire population even if the host population satisfies the ESS condition. In analysing finite population, one has to consider a microscopic birth-death process and also fitness values which are slightly different from that in the corresponding infinite case. One of the paradigmatic finite population dynamics is formulated as a Moran process with frequency-dependent fitness. 

Moran process~\cite{Moran1958} is one of the simplest stochastic birth-death process in a finite population. In Moran process, one considers a finite generation-wise overlapping population consisting of $N$ individuals. The individual are of two types, which putting in the context of the present paper, can be taken to be $E_{n-1}(p,q)$ and ALLD. Now, one individual is chosen randomly for reproduction with a probability proportional to its fitness and allowed reproduce. Subsequently, a reproduced individual replaces any single individual chosen at random so that the population size remains temporally constant at $N$. Moran process has two absorbing states: All ALLD individuals and all $E_{n-1}(p,q)$ individuals. As mentioned in Sec.~\ref{sec:intro}, the relevant concept---analogous to ESS---in this case is that of ESS$_N$ that is defined through fixation probability. The fixation probability of, say, ALLD is the probability with which a single ALLD individual---actually, its descendants---can take over the rest of the population ($N-1$ $E_{n-1}(p,q)$ individuals). 

The fitnesses of ALLD and $E_{n-1}(p,q)$ individuals can be expressed~\cite{2013_N} respectively as $f^A_i=1-w+w F^A_i$ and $f^E_i=1-w+w F^E_i$ respectively where $w$ is the strength of the game's (or differential selection's) contribution to the fitnesses; the parameter $0\leq w\ll 1$ implies weak selection. $F^A_i$ and $F^E_i$ are respectively found~\cite{Kurokawa_2009, dga} to be (see \autoref{tab:payoffm}):
\begin{subequations}
\begin{eqnarray}
&&F^A_i=\sum_{k=1}^n \left[\frac{{}^{i-1}C_{n-k}\, {}^{N-i}C_{k-1}}{{}^{N-1}C_{n-1}}\right] a_k,~~{\rm and}\\
&&F^E_i=\sum_{k=1}^n\left[ \frac{{}^{i}C_{n-k}\, {}^{N-i-1}C_{k-1}}{{}^{N-1}C_{n-1}}\right] b_k,
\end{eqnarray} 
\end{subequations}
where, $i\in\{0,\,1,\,\cdots,\, N\}$ is the number of ALLD individuals in the population. In the limit of weak selection, the conditions for  $E_{n-1}(p,q)$ to be ESS$_N$ can be written down using the results~\cite{Kurokawa_2009} available in the literature:
\begin{subequations}
\begin{eqnarray}
&&(N-1)a_n < (N-n)b_n + (n-1)b_{n-1}~~{\rm and}\qquad\qquad\quad
\label{eq:condition1}\\
&&\sum_{k=1}^{n} k(a_k -b_k) N < -n^2 b_n + \sum_{k=1}^{n-1}kb_k+ \sum_{k=1}^n (n+1-k)a_k.\nonumber\\
\label{eq:condition2}
\end{eqnarray}
\end{subequations}
If inequality~(\ref{eq:condition1}) is satisfied, selection opposes a single ALLD individual invading a finite population of $E_{n-1}(p, q)$. On the other hand, if inequality~(\ref{eq:condition2}) is satisfied, then the selection opposes ALLD from replacing $E_{n-1}(p, q)$ which means that the fixation probability of ALLD is less than $1/N$. We now want to find all the values of $p$ and $q$ for which $E_{n-1}(p,q)$ is ESS$_N$ against ALLD.

%%%%%%%%%%%%%%%%%%%%%%%%%%%%%%%%%%%%
%%%%%%%%%%%%%%%%%%%%%%%%%%%%%%%%%%%%%%%%%%

%%%%%%%%%%%%%%%%%%%%%%%%%%%%%%%%%%%%%%%

\subsection{Large population limit}
There are quite a few parameter, viz., $b$, $c$, $\delta$, $n$ and $N$, to ponder about. In order to systematically understand their effects, it is pragmatic to first work in the limit of large population (i.e., large $N$) because then conditions~(\ref{eq:condition1})~and~(\ref{eq:condition2}) become $N$ independent:
\begin{subequations}
\begin{eqnarray}
&&a_n<b_n,
\label{eq:cond1}\\
&&\sum_{k=1}^n k (a_k-b_k) < 0.
\label{eq:cond2}
\end{eqnarray}
\end{subequations}
Furthermore, while we take different $b$ and $c$ in our numerical investigations, we work with two special classes---one with $b/c$ fixed and the other with $b/nc$ fixed. The factor $b/c$, the benefit-to-cost ratio, is a known~\cite{BoydBook2007} important quantity that arises in the study of effect of reciprocity in the emergence of cooperation; it is quite natural to measure benefit relative to the cost paid because the benefit for per unit cost is what one would want to maximise. In totally different context, benefit-cost ratio~\cite{2010_Levin} is a standard indicator used in cost-benefit analysis in economics. One can intuit that in multiplayer games, because group-size is another factor to consider, benefit-cost ratio per individual in a group should also be of interest. Further justification is in the interesting results we obtain later. Next, as done in the case of ESS in Sec.~\ref{sec:ess}, we find all the values of $p$ and $q$ for which $E_{n-1}(p,q)$ is ESS$_N$ against ALLD for some minimal $\delta=\delta_{\rm min}$. All the results presented henceforth are obtained by uniformly varying $p$ and $q$ from 0 to 1 in the step size of 0.02; and $\delta_{\rm min}$ is found among all values of $\delta\in(0,1)$ with resolution of $0.01$.

\subsubsection{Fixed $b/c$ and effect of group size $n$}
Let us focus on Fig.~\ref{fig:b_by_c}. First in the first row we fix the $b/c$ to $1.5$ by choosing $b=1.5$ and $c=1$ and see the effect of interaction group size $n$ by taking the values $n=2,~3,~4,~{\rm and}~20$. The region on the parameter space shrinks where the strategy $E_{n-1}(p,q)$ is evolutionarily stable against ALLD as shown in Fig.~\ref{fig:b_by_c}(a)--Fig.~\ref{fig:b_by_c}(d). In each of this plot one point is common: For a given value of generosity ($q\leq q_{\rm max}$), the value of $\delta_{\rm min}$ increases with the increase in non-reciprocity, and beyond a maximum value of non-reciprocity $E_{n-1}(p,q)$ ceases to be ESS$_N$; similarly, for a given value of non-reciprocity ($1-p\le 1-p_{\rm min}$), the value of $\delta_{\rm min}$ increases with increase in generosity, and beyond a maximum value of generosity $E_{n-1}(p,q)$ ceases to be ESS$_N$. 

Interestingly, with increase in the group size, the minimum required discount factor $\delta_{\rm min}$ increases with $n$ for every $E_{n-1}(p,q)$ that is ESS$_N$ throughout. In other words, it implies that effective game length needs to be longer for an $E_{n-1}(p,q)$ to be evolutionarily stable as group size increases. It is in line with similar results reported in~\cite{Boyd1988}. Next, in later rows of Fig.~\ref{fig:b_by_c}, we use different $b$ and $c$ but keep their ratio fixed to $1.5$ to check the robustness of the effect of $b/c$. We find that the effect of group size is independent of independent specific values of $b$ and $c$ as long as $b/c$ is same. Finally, we observe decrease in the values of $R_1$ and $R_2$ with increase in interaction group size. This implies, as visual inspection of the figure suggests as well, that the sustenance of cooperation becomes relatively easier for generous reactive strategies compared to the non-reciprocal reactive strategies as $n$ increases; for large enough $n$, only TFT$_{n-1}$ is expected to be evolutionarily stable.

\subsubsection{Fixed $b/nc$ and effect of group size $n$}
Now, let us inspect Fig.~\ref{fig:b_by_nc}. Throughout the subplots of the figure, we keep $b/nc$ fixed at 0.75. Row-wise $b$ increases and column-wise $n$ increases. Thus, we immediately note that along any column---because $n$ and $b/c$ is constant---all the three plots are identical. Along any row, $b/c$ and $n$ increase together and we note that $E_{n-1}(p,q)$ is ESS$_N$ for rather more values of $p$ and $q$; both $q_{\rm max}$ and $1-p_{\rm min}$ increase. In other words, more generous and more non-reciprocal reactive strategies tend to become evolutionarily stable as the group of simultaneously interacting players becomes bigger. We furthermore observe that for fixed $b/nc$, at the same $p$ and $q$, the minimum discount factor $\delta_{\rm min}$ becomes lesser than what is required in the case of smaller group sizes. In other words, the expected game length decreases for larger group size when $b/nc$ is constant and therefore it becomes easier for the strategy $E_{n-1}(p,q)$ to become evolutionarily stable. 

What is most interesting is that for fixed $b/nc$, $R_1\approx R_2\approx 0.5$. This means that the maximum extent of generosity that an evolutionary stable finite population consisting of completely reciprocal individuals can sustain is equal to the maximum extent of non-reciprocity that an evolutionary stable finite population consisting of completely non-generous individuals can sustain. This fact and its comparison with the case of infinite population is presented in Fig.~\ref{fig:asymmetry} where we clearly observe that constant $b/nc$ makes generosity and non-reciprocity symmetric in their effect on each other but for fixed $b/c$ they become more and more asymmetric with the increase in the group size.

\subsection{Effect of population size $N$}
Qualitatively, we expect the conclusions discussed for the case of large $N$ limit to show up even for the case of smaller $N$ values. In order to avoid trivial repetition sans any new insight we do not explicitly present those results for the smaller $N$ values. However, for the sake of completeness, we mention how change in the value of $N$ itself can effect the non-reciprocity--generosity symmetry for the case of more than two players simultaneously interacting. We recall that we must now use conditions~(\ref{eq:condition1})~and~(\ref{eq:condition2}) instead of 
conditions~(\ref{eq:cond1})~and~(\ref{eq:cond2}) to find if an $E_{n-1}(p,q)$ is ESS$_N$.

In Fig.~\ref{fig:finiteN}, we fix $n=3$ for illustrative purpose. Along the rows $N$ varies and along the columns $b/c$ is varied. For any $N$, increase in the benefit to cost ratio enhances the set of $E_{n-1}(p,q)$ that can become ESS$_N$ and simultaneously make $R_1$ and $R_2$ approach $0.5$ from the lower side. Also, smaller expected game length suffices for making an $E_{n-1}(p,q)$ ESS$_N$. For a fixed $b/c$, increase in the population size interestingly has exactly same effect: For any $b/c$, increase in $N$ increase the number of values of $p$ and $q$ such that $E_{n-1}(p,q)$ can become ESS$_N$ and simultaneously $R_1$ and $R_2$ monotonically approach $0.5$ from the smaller values. Here also, for making an $E_{n-1}(p,q)$ ESS$_N$, one requires smaller expected game length as $N$ increases.

Hence, for a given group size, the establishment of cooperation is easier in a larger population and non-reciprocity--generosity symmetry is more probable in a larger population. On the other hand, for large enough population, we recall from earlier discussions that the evolutionary stability of $E_{n-1}(p,q)$ and the non-reciprocity--generosity symmetry become easier for small-sized interacting groups. We find this subtle interplay between the interaction group size and the population size on the evolutionary stability of reactive strategies quite interesting.

\section{Conclusion and Discussion}
\label{sec:discussion}
Summarizing, we have used the concepts of ESS and ESS$_N$ to study the evolutionary stability of multiplayer reactive strategies, $E_{n-1}(p,q)$, amid the threat by ALLD mutants in multiplayer infinitely repeated prisoner's dilemma game. In the light of the fact that a completely non-generous and reciprocal strategy (i.e., TFT$_{n-1}$) is evolutionary stable both in finite and infinite populations for large enough expected game length, it is a natural question to as how much generous or non-reciprocal the individuals of the host population can become while not giving way to the always defecting mutants. This paper answers this question and additionally highlights the relative efficacies of being generous and being non-reciprocal in establishing cooperation, even if in coexistence with some amount of defection. 

Specifically, we have found that for a given benefit to cost ratio, increasing interaction group size skews the non-reciprocity--generosity asymmetry towards generosity. However, for a given benefit to cost ratio per player in the group, the effect of non-reciprocity and generosity are symmetric as far as their robustness against mutant invasion is concerned; increasing group size, however, allows for increase in both generous and non-reciprocal strategies that are evolutionary stable. Finally, we have highlighted the interplay between the finiteness of the population and the interaction group size in the context of the main question of this paper. It appears that the emergence of cooperation is comparatively more likely to emerge in larger populations where the interaction group size is smaller and the non-reciprocity--generosity asymmetry is minimal.

Generosity is very much desirable in a population as it has an important characteristic of forgiving some defection, which might have been played erroneously by the opponent, and thus maintain higher payoff in such a situation. In fact we established in this paper that whenever possible, generous strategies fare better than non-reciprocal strategies in establishing cooperation in multiplayer interactions. We remind the readers that there exists a much bigger set of reactive strategies left to be explored, viz., $E_m(p,q)$ with $0\le m<n-1$; this paper has concentrated only on $E_{n-1}(p,q)$, the hardest of them all. The softer reactive strategies for a fixed $q$ may also be interpreted~\cite{Kurokawa2010} as relatively more generous. How non-reciprocity compares against this generosity is something worth looking into. Moreover, how two strategies ---$E_m(p,q)$ and $E_{m'}(p,q)$ ($m\ne m'$)---fare against each other in the context of non-reciprocity--generosity asymmetry, is also an interesting aspect worth investigating.  

Furthermore, in future, we could leave the setup of homogeneously strategied population and explore the extensions of our result in the heterogeneous population~\cite{Nowak1992}. Of course, going beyond the formalism of the reactive strategies is analytically intractable but it would be exciting to numerically investigate the effect of past memory of arbitrary length on the non-reciprocity--generosity asymmetry. Last but not the least, the assumption of idealised unstructured population should also be relaxed for making more real-world connection and invoking some network topologies~\cite{2012Paulo, 2012Julia} %a classic book or a recent review} 
to model interaction structure among individuals of the structured population with multiplayer games would be a potential future direction of extension.
\acknowledgements
The authors are thankful to Arunava Patra for verifying some of the calculations presented in this paper.
%%%%%%%%%% Insert bibliography here %%%%%%%%%%%%%%
\bibliographystyle{apsrev4-1}
\bibliography{Sadhukhan_etal_manuscript.bib}

\end{document}